\newcommand{\diw}{\ensuremath{\delta_{i}(w)}}

\newcommand{\ra}{\ensuremath{\rightarrow}}

\newcommand{\dit}{\ensuremath{\delta_{i}^{t}}}
\newcommand{\ditw}{\ensuremath{\dit (w)}}

\newcommand{\edit}{\ensuremath{e(\dit)}}
\newcommand{\Dit}{\ensuremath{\Delta_{i}^{t}}}
\newcommand{\Ditt}{\ensuremath{\Delta_{i}^{t+1}}}
\newcommand{\Ditw}{\ensuremath{\Dit(w)}}

\newcommand{\ditt}{\ensuremath{\delta_{i}^{t+1}}}
\newcommand{\dittw}{\ensuremath{\ditt (w)}}

\newcommand{\djtt}{\ensuremath{\delta_{j}^{t+1}}}
\newcommand{\djttw}{\ensuremath{\djtt (w)}}
\newcommand{\editt}{\ensuremath{e(\ditt)}}
\newcommand{\edjtt}{\ensuremath{e(\djtt)}}

\newcommand{\djt}{\ensuremath{\delta_{j}^{t}}}
\newcommand{\djtw}{\ensuremath{\delta_{j}^{t}(w)}}
\newcommand{\edjt}{\ensuremath{e(\djt)}}

\newcommand{\Dittw}{\ensuremath{\Ditt (w)}}

\newcommand{\Djt}{\ensuremath{\Delta_{j}^{t}}}
\newcommand{\Djtw}{\ensuremath{\Djt (w)}}

\newcommand{\Iji}{\ensuremath{I_{ji}}}
\newcommand{\Iij}{\ensuremath{I_{ij}}}

\newcommand{\Pro}{\ensuremath{\mathbf{Pr}}}
\newcommand{\D}{\ensuremath{\mathcal{D}}}
\newcommand{\Dw}{\ensuremath{\mathcal{D}(w)}}

\newcommand{\forallb}{\ensuremath{\mathbf{\forall}}}

\newcommand{\charts}[1]{\includegraphics*[width=2.5in]{#1}}

\documentclass{article}
\usepackage{natbib,amsmath,psfrag,graphicx, subfigure, url, fancyhdr,hyperref}
\usepackage[all]{xy}

\hypersetup{bookmarksopen,
  bookmarksnumbered, 
  pdftitle={Predicting the Expected Behavior of Agents that Learn About Agents: The CLRI Framework},
  pdfauthor={Jose M. Vidal and Edmund H. Durfee},
  pdfsubject={Learning in MASs},
  pdfkeywords={Learning, MAS}, 
  pdfview=FitH,
  pdfstartview=FitH 
}

\title{Predicting the Expected Behavior of Agents that Learn About
  Agents: The CLRI Framework}

\author{Jos\'{e} M. Vidal and Edmund H. Durfee \\
Swearingen Engineering Center, University of South \\
Carolina, Columbia, SC, 29208 \\
Advanced Technology Laboratory, University of \\
  Michigan, Ann Arbor, MI, 48102}
\begin{document}
\maketitle
\begin{abstract}
  We describe a framework and equations used to model and predict the
  behavior of multi-agent systems (MASs) with learning agents. A
  difference equation is used for calculating the progression of an
  agent's error in its decision function, thereby telling us how the
  agent is expected to fare in the MAS. The equation relies on
  parameters which capture the agent's learning abilities, such as its
  change rate, learning rate and retention rate, as well as relevant
  aspects of the MAS such as the impact that agents have on each
  other. We validate the framework with experimental results using
  reinforcement learning agents in a market system, as well as with
  other experimental results gathered from the AI literature. Finally,
  we use PAC-theory to show how to calculate bounds on the values of
  the learning parameters.
\end{abstract}
  Multi-Agent Systems, Machine Learning, Complex Systems.

\thispagestyle{fancy}
\lhead{}
\chead{Autonomous Agents and Multiagent Systems, January, 2003.}
\cfoot{\copyright{} Kluwer Academic Publishers 2002.}

\section{Introduction}
\label{sec:Introduction}

With the steady increase in the number of multi-agent systems (MASs)
with learning agents \cite{durfee:97,chavez:96,etzioni:96,stone97a}
the analysis of these systems is becoming increasingly important.
Some of the research in this area consists of experiments where a
number of learning agents are placed in a MAS, then different learning
or system parameters are varied and the results are gathered and
analyzed in an effort to determine how changes in the individual agent
behaviors will affect the system behavior.  We have learned about the
dynamics of market-based MASs using this approach \cite{vidal:98b}.
However, in this article we will take a step beyond these
observation-based experimental results and describe a framework that
can be used to model and predict the behavior of MASs with learning
agents.  We give a difference equation that can be used to calculate
the progression of an agent's error in its decision function.  The
equation relies on the values of parameters which capture the agents'
learning abilities and the relevant aspects of the MAS. We validate
the framework by comparing its predictions with our own experimental
results and with experimental results gathered from the AI literature.
Finally, we show how to use probably approximately correct (PAC)
theory to get bounds on the values of some of the parameters.

\begin{figure}
  \begin{center}
    \psfrag{d1}{$\delta_1$}
    \psfrag{d2}{$\delta_2$}
    \psfrag{d3}{$\delta_3$}
    \psfrag{d}{{\tiny $\delta$}}
    \includegraphics{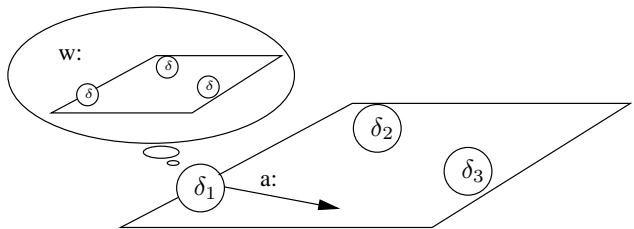}
    \caption{The agents in a MAS.}
    \label{fig:problem}
  \end{center}
\end{figure}

The types of MAS we study are exemplified by the abstract
representation shown in Figure~\ref{fig:problem}. We assume that the
agents observe the physical state of the world (denoted by $w$ in the
figure) and take some action ($a$) based on their observation of the
world state. An agent's mapping from states to actions is denoted by
the decision function ($\delta$) inside the agent.  Notice that the
``physical'' state of the world includes everything that is directly
observable by the agent using its sensors. It could include facts such
as a robot's position, or the set outstanding bids in an auction,
depending on the domain. The agent does not know the decision
functions of the other agents.  After taking action an agent can
change its decision function as prescribed by whatever
machine-learning algorithm the agent is using.

We have a situation where agents are changing their decision function
based on the effectiveness of their actions. However, the
effectiveness of their actions depends on the other agents' decision
functions.  This scenario leads to the immediate problem: if all the
agents are changing their decisions functions then it is not clear
what will happen to the system as a whole. Will the system settle to
some sort of equilibrium where all agents stop changing their decision
functions? How long will it take to converge? How do the agents'
learning abilities influence the system's behavior and possible
convergence? These are some of the questions we address in this
article.

Section~\ref{sec:Fram-Model-MASs} presents our framework for
describing an agent's learning abilities and the error in its
behavior.  Section~\ref{sec:Calc-Agent's-Error} presents an equation
that can be used to predict an agent's expected error, as a function
of time, when the agent is in a MAS composed of other learning agents.
This equation is simplified in Section~\ref{sec:Assum-Cond-Indep} by
making some assumptions about the type of MAS being modeled.
Section~\ref{sec:Volatility} then defines the last few parameters used
in our framework---volatility and impact.
Section~\ref{sec:Example-with-2} gives an illustrative example of the
use of the framework.  The predictions made by our framework are
verified by our own experiments, as shown in
Section~\ref{sec:Simple-Application}, and with the experiments of
others, as shown in Section~\ref{sec:Appl-our-Theory}. The use of PAC
theory for determining bounds on the learning parameters is detailed
in Section~\ref{sec:Bound-Learn-Rate}. Finally
Section~\ref{sec:Related-Work} describes some of the related work and
Section~\ref{sec:Summary} summarizes our claims.

\section{A Framework for Modeling MASs}
\label{sec:Fram-Model-MASs}

In order to analyze the behaviors of agents in MASs composed of
learning agents, we must first construct a formal framework for
describing these agents. The framework must state any assumptions and
simplifications it makes about the world, the agents, and the agents'
behaviors. It must also be mathematically precise, so as to allow us
to make quantitative predictions about expected behaviors. Finally,
the simplifications brought about because of the need for mathematical
precision should not be so constraining that they prevent the
applicability of the framework to a wide variety of learning
algorithms and different types of MASs.  We now describe our framework
and explain the types of MASs and learning behaviors that it can
capture.

\subsection{The World and its Agents}
\label{sec:World-its-Agents}

A MAS consists of a finite number of agents, actions, and world
states. We let $N$ denote the finite set of agents in the system.  $W$
denotes the finite set of world states.  Each agent is assumed to have
a set of perceptors (e.g., a camera, microphone, bid queue) with which
it can perceive the world. An agent uses is sensors to ``look'' at the
world and determine which world state $w$ it is in; the set of all
these states is $W$.  $A_i$, where $|A_i| \geq 2$, denotes the finite
set of actions agent $i \in N$ can take.

We assume discrete time, indexed in the various functions by the
superscript $t$, where $t$ is an integer greater than or equal to 0.
The assumption of discrete time is made, for practical reasons, by a
number of learning algorithms. It means that, while the world might be
continuous, the agents perceive and learn in separate discrete
time steps.

We also assume that there is only one world state $w$ at each time,
which all the agents can perceive in its completeness. That is, we
asume the enviroment is accessible (as defined in
\cite[p46]{ai:modern:approach}).  This assumption holds for market
systems in which all the actions of all the agents are perceived by
all the agents, and for software agent domains in which all the agents
have access to the same information. However, it might not hold for
robotic domains where one agent's view of the world might be obscured
by some physical obstacle. Even in such domains, it is possible that
there is a strong correlation between the states perceived by each
agent. These correlations could be used to create equivalency classes
over the agents' perceived states, and these classes could then be
used as the states in $W$.

Finally, we assume the environment is determistic
\cite[p46]{ai:modern:approach}. That is, the agents' combined actions
will always have the expected effect. Of course, agent $i$ might not
know what action agent $j$ will take so $i$ might not know the
eventual effect of its own individual action.

\subsection{A Description of Agent Behavior}
\label{sec:Agents}

In the types of MASs we are modeling, every agent $i$ perceives the
state of the world $w$ and takes an action $a_i$, at each time step.
We assume that every agent's \textbf{behavior}, at each moment in
time, can be described with a simple state-to-action mapping. That is,
an agent's choice of action is solely determined by its current
state-to-action mapping and the current world $w$.

Formally, we say that agent $i$'s behavior is represented by a
\textbf{decision function} (also known as a ``policy'' in control
theory and a ``strategy'' in game theory), given by $\delta_{i}^t:W
\ra A_i$. This function maps each state $w \in W$ to the action $a_i
\in A_i$ that agent $i$ will take in that state, at time $t$. This
function can effectively describe any agent that deterministically
chooses its action based on the state of the world. Notice that the
decision function is indexed with the time $t$. This allows us to
represent agents that change their behavior.

The action agent $i$ \emph{should} take in each state $w$ is given by
the \textbf{target function} $\Delta_{i}^t:W \ra A_i$, which also maps
each state $w \in W$ to an action $a_i \in A_i$. The agent does not
have direct access to its target function. The target function is used
to determine how well an agent is doing. That is, it represents the
``perfect'' behavior for a given agent. An agent's learning task is to
get its decision function to match its target function as much as
possible.

Since the choice of action for agent $i$ often depends on the actions
of other agents, the target function for $i$ needs to take these
actions into account. That is, in order to generate \Dit{}, one would
need to know \djtw{} for all $j \in N_{-i}$ and $w \in W$.  These
\djtw{} functions tell us the actions that all the other agents will
take in every state $w$.  For example, in order for one to determine
what an agent should bid in every world $w$ of an auction-based market
system, one will need to know what the other agents will bid in every
world $w$. One can use these actions, along with the state $w$, in
order to identify the best action for $i$ to take.

\begin{figure}
  \centerline{ 
    \xymatrix{ *\txt{New world $w^t \in \D$} \ar[r] & *\txt{Perceive world $w^t$} 
      \ar[r] & *\txt{Take action $\delta_i(w^t)$}\ar[d] \\
      & *\txt{Learn}\ar[lu]_{t \leftarrow t + 1} & *\txt{Receive payoff\\ or feedback.}\ar[l]
      }}
    \caption{Action/Learn loop for an agent.}
    \label{fig:action-learn}
\end{figure}
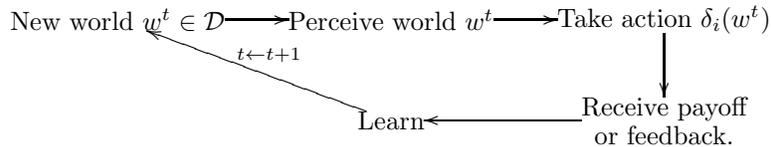

An agent's \ditw{} can change over time, so that $\ditt \neq \dit$.
These changes in an agent's decision function reflect its learned
knowledge. The agents in the MASs we consider are engaged in the
discrete action/learn loop shown in Figure~\ref{fig:action-learn}. The
loop works as follows: At time $t$ the agents perceive a world $w^t
\in W$ which is drawn from a fixed distribution \Dw{}.  They then each
take the action dictated by their \dit{} functions; all of these
actions are assumed to be taken effectively in parallel. Lastly, they
each receive a payoff which their respective learning algorithms use
to change the \dit{} so as to, hopefully, better match \Dit{}.  By
time $t+1$, the agents have new \ditt{} functions and are ready to
perceive the world again and repeat the loop.  Notice that, at time
$t$, an agent's \Dit{} is derived by taking into account the \djt{} of
all other agents $j \in N_{-i}$.

We assume that \Dw{} is a fixed probability distribution from which we
take the worlds seen at each time. This assumption is not unreasonably
limiting.  For example, in an economic domain where the new state is
the new good being offered, or in an episodic domain where the agents
repeatedly engage in different games (e.g. a Prisoner's Dilemma
competition) there is no correlation between successive world states
or between these states and the agents' previous actions.  However, in
a robotic domain one could argue that the new state of the world will
depend on the current state of the world; after all, the agents
probably move very little each time step.

Our measure of the correctness of an agent's behavior is given by our
\textbf{error} measure. We define the error of agent $i$'s decision
function \ditw{} as
\begin{equation}
  \label{eq:error}
  \begin{split}
    \edit &= \sum_{w \in W} \Dw \Pro[\ditw \neq \Ditw] \\
    &= \Pro_{w \in \D} [\ditw \neq \Ditw].
  \end{split}
\end{equation}

\edit{} gives us the probability that agent $i$ will take an incorrect
action; it is in keeping with the error definition used in
computational learning theory \cite{intro:clt}.  We use it to gauge
how well agent $i$ is performing. An error of 0 means that the agent
is taking all the actions dictated by its target function. An error of
1 means that the agent never takes an action as dictated by its target
function. Each action the agent takes is either correct or incorrect,
that is, it either matches the target function or it does not. We do
not model degrees of incorrectness. However, since the error is
defined as the average over all possible world states, an agent that
takes the correct action in most world states will have a small error.
Extending the theory to handle degrees of incorrectness is one of
the subjects of your continuing work, see
Section~\ref{sec:future-work}. All the notation from this section is
summarized in Figure~\ref{fig:summary-notation}.

\begin{figure}
  \begin{center}
    \fbox{ \parbox{4.5in}{
        \begin{description}
        \item[$N$] the set of all agents, where $i \in N$ is one
          particular agent.
        \item[$W$] the set of possible states of the world, where $w
          \in W$ is one particular state.
        \item[$A_i$] the set of all actions that agent $i$ can take.
        \item[$\delta_i^t: W \ra A_i$] the \textbf{decision} function
          for agent $i$ at time $t$. It tells which action agent $i$
          will take in each world.
        \item[$\Delta_i^t: W \ra A_i$] the \textbf{target} function
          for agent $i$ at time $t$. It tells us what action agent $i$
          should take. It takes into account the actions that other
          agents will take.
        \item[$e(\delta_i^t)$] $= \Pro[\delta_i^t(w) \neq
          \Delta_i^t(w) \,|\, w \in \D]$ the \textbf{error} of agent
          $i$ at time $t$. It is the probability that $i$ will take an
          incorrect action, given that the worlds $w$ are taken from
          the fixed probability distribution \D.
        \end{description}
        } }
    \caption{Summary of notation used for describing a MAS and the agents in it.}
    \label{fig:summary-notation}
  \end{center}
\end{figure}

\subsection{The Moving Target Function Problem}
\label{sec:Moving-Targ-Funct}

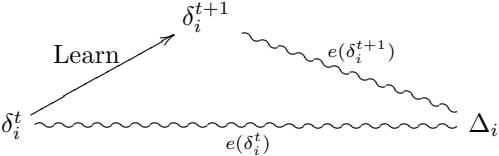
\begin {figure}[thbp]
  \centerline{
    \xymatrix {
      & &  \ditt{}  \ar@{~}[rrrd]^{e(\ditt)}
      & & &  \\
      \dit{} \ar[rru]^{\txt{Learn}} \ar@{~}[rrrrr]_{e(\dit)} &&
      & & &  \Delta_i
      } }
  \caption{The traditional learning problem.}
  \label{fig:trad-learn}
\end{figure}

The learning problem the agent faces is to change its \ditw{} so that
it matches \Ditw{}.  If we imagine the space of all possible decision
functions, then agent $i$'s \dit{} and \Dit{} will be two points in
this space, as shown in Figure~\ref{fig:trad-learn}.  The agent's
learning problem can then be re-stated as the problem of moving its
decision function as close as possible to its target function, where
the distance between the two functions is given by the error \edit{}.
This is the traditional machine learning problem.

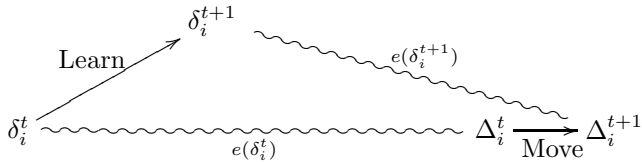
\begin{figure}[thbp]
  \centerline{
    \xymatrix{
      & &  \ditt{}  \ar@{~}[rrrrd]^{e(\ditt)}
      & & & \\
      \dit{} \ar[rru]^{\txt{Learn}} \ar@{~}[rrrrr]_{e(\dit)} &&
      & & & \Dit \ar[r]_{\txt{Move}} & \Ditt
      }}
  \caption{The learning problem in learning MASs.}
  \label{fig:learn-mas}
\end{figure}

However, once agents start to change their decision functions (i.e.,
change their behaviors) the problem of learning becomes more
complicated because these changes might cause changes in the other
agents' target functions. We end up with a moving target function, as
seen in Figure~\ref{fig:learn-mas}. In these systems, it is not clear
if the error will ever reach 0 or, more generally, what the expected
error will be as time goes to infinity. Determining what will happen
to an agent's error in such a system is what we call the
\textbf{moving target function problem}, which we address in this
article. However, we will first need to define some parameters that
describe the capabilities of an agent's learning algorithm.

\subsection{A Model of Learning Algorithms}
\label{sec:Model-Learn-Algor}

An agent's learning algorithm is responsible for changing \dit{} into
\ditt{} so that it is a better match of \Dit{}. Different machine
learning algorithms will achieve this match with different degrees of
success.  We have found a set of parameters that can be used to model
the effects of a wide range of learning algorithms. The parameter are:
Change rate, Learning rate, Retention rate, and Impact; and they will
be explained in this section, except for Impact which will be
introduced in Section~\ref{sec:Volatility}. These parameters, along
with the equations we provide, form the \textbf{CLRI} framework (the
letters correspond to the first letter of the parameters' names).

After agent $i$ takes an action and receives some payoff, it activates
its learning algorithm, as we showed in Figure~\ref{fig:action-learn}.
The learning algorithm is responsible for using this payoff in order
to change \dit{} into \ditt{}, making \ditt{} match \Dit{} as much as
possible. We can expect that for some $w$ it was true that $\ditw =
\Ditw$, while for some other $w$ this was not the case. That is,
some of the $w \ra a_i$ mappings given by \ditw{} might have been
incorrect.  In general, a learning algorithm might affect both the
correct and incorrect mappings. We will treat these two cases
separately.

We start by considering the incorrect mappings and define the
\textbf{change rate} of the agent as the probability that the agent
will change at least one of its incorrect mappings. Formally, we
define the change rate $c_i$ for agent $i$ as
\begin{equation}
  \label{eq:54} 
  \forallb_{w} \; \Pro[\dittw \neq \ditw \, |\, \ditw \neq \Ditw] = c_i.
\end{equation}

The change rate tells us the likelihood of the agent changing an
incorrect mapping into something else. This ``something else'' might
be the correct action, but it could also be another incorrect action.
The probability that the agent changes an incorrect mapping to the
correct action is called the \textbf{learning rate} of the agent.
It is defined as $l_i$ where
\begin{equation}
  \label{eq:51}
  \forallb_{w} \; \Pro[\dittw = \Ditw \, | \, \ditw \neq
  \Ditw] = l_{i}.
\end{equation}
There are two constraints which must always be satisfied by these two
rates. Since changing to the correct mapping implies that a change was
made, the value of $l_i$ must be less than or equal to $c_i$, that is,
$l_i \leq c_i$ must always be true.  Also, if $|A_i| = 2$ then $c_i =
l_i$ since there are only two actions available, so the one that is
not wrong must be right. 

The complementary value for the learning rate is $1- l_i$ and refers
to the probability that an incorrect mapping does not get changed to a
correct one.  An example learning rate of $l_i = .5$ means that if
agent $i$ initially has all mappings wrong it will make half of them
match the original target function after the first iteration.

We now consider the agent's correct mappings and define the
\textbf{retention rate} as the probability that a correct mapping will
stay correct in the next iteration. The retention rate is given by
$r_i$ where 
\begin{equation}
  \label{eq:52}
  \forallb_{w} \; \Pro[\dittw = \Ditw \, | \, \ditw = \Ditw].
 = r_i.
\end{equation}
We propose that the behavior of a wide variety of learning algorithms
can be captured (or at least approximated) using appropriate values
for $c_i$, $l_i$, and $r_i$. Notice, however, that these three rates
claim that the $w \ra a$ mappings that change are independent of the
$w$ that was just seen. We can justify this independence by noting
that most learning algorithms usually perform some form of
generalization.  That is, after observing one world state $w$ and the
payoff associated with it, a typical learning algorithm is able to
generalize what it learned to some other world states. This
generalization is reflected in the fact that the change, learning, and
retention rates apply to all $w$'s. However, a more precise model
would capture the fact that, in some learning algorithms, the mapping
for the world state that was just seen is more likely to change than
the mapping for any other world state.

The rates are not time dependent because we assume that agents use one
learning algorithm during their lifetimes. The rates capture the
capabilities of this learning algorithm and, therefore, do not need to
vary over time. 

Finally, we define \textbf{volatility} to mean the probability that
the target function will change from time $t$ to time $t+1$. Formally,
volatility is given by $v_i$ where
\begin{equation}
  \label{eq:53}
  \forallb_{w} \; \Pro[\Dittw \neq \Ditw] = v_i
\end{equation}
In Section~\ref{sec:Volatility}, we will show how to calculate $v_i$
in terms of the error of the other agents. We will then see that
volatility is not a constant but, instead, varies with time.

\section{Calculating the Agent's Error}
\label{sec:Calc-Agent's-Error}

We now wish to write a difference equation that will let us calculate
the agent's expected error, as defined in Eq.~\eqref{eq:error}, at
time $t+1$ given the error at time $t$ and the other parameters we
have introduced. We can do this by observing that there are two
conditions that determine the new error: whether $\Dittw{} = \Ditw{}$
or not, and whether $\ditw{} = \Ditw{}$ or not.  If we define $a
\equiv \Dittw{} = \Ditw{}$, and $b \equiv \ditw{} = \Ditw{}$, we can
then say that we need to consider the four cases where: $a \wedge b$,
$a \wedge \neg b$, $\neg a \wedge b$, and $\neg a \wedge \neg
b$. Formally, this implies that
\begin{equation}
  \label{eq:1}
  \begin{split}
    \Pro&[\dittw \neq \Dittw] = \\
    & \Pro[\dittw \neq \Dittw \wedge a \wedge b] + 
    \Pro[\dittw \neq \Dittw \wedge a \wedge \neg b] + \\
    & \Pro[\dittw \neq \Dittw \wedge \neg a \wedge b] + 
    \Pro[\dittw \neq \Dittw \wedge \neg a \wedge \neg b],
  \end{split}
\end{equation}
since the four cases are exclusive of each other. Applying the chain
rule of probability, we can rewrite each of the four terms in order to
get
\begin{equation}
  \label{eq:2}
  \begin{split}
    \Pro[\dittw \neq \Dittw] &= \Pro[a \wedge b] \cdot \Pro[\dittw \neq \Dittw \,|\, a\wedge b] + \\
    & \Pro[a \wedge \neg b] \cdot \Pro[\dittw \neq \Dittw \,|\, a\wedge
  \neg b] + \\
  & \Pro[\neg a \wedge b] \cdot \Pro[\dittw \neq \Dittw \,|\, \neg
  a\wedge  b] + \\
  & \Pro[\neg a \wedge \neg b] \cdot \Pro[\dittw \neq \Dittw \,|\, \neg
  a\wedge \neg b].
\end{split}
\end{equation}
We can now find values for these conditional probabilities. We start
with the first term where, after replacing the values of $a$ and $b$,
we find that
\begin{equation}
  \label{eq:3}
  \Pro[\dittw \neq \Dittw \,|\, \Dittw = \Ditw \wedge \ditw = \Ditw] = 1 - r_i.
\end{equation}
Since the target function does not change from time $t$ to $t+1$ and
the agent was correct at time $t$, the agent will also be correct at
time $t+1$; \emph{unless} it changes its correct $w \ra a$ mapping. The 
agent changes this mapping with probability $1 - r_i$. 

The value for the second conditional probability is
\begin{equation}
  \label{eq:4}
  \Pro[\dittw \neq \Dittw \,|\, \Dittw = \Ditw \wedge \ditw \neq \Ditw] = 1 - l_i.
\end{equation}
In this case the target function still stays the same but the agent
was incorrect. If the agent was incorrect then it will change its
decision function to match the target function with probability
$l_i$. Therefore, the probability that it will be incorrect next time
is the probability that it does not make this change, or $1 -
l_i$. 

The third probability has a value of
\begin{equation}
  \label{eq:5}
  \begin{split}
  \Pro&[\dittw \neq \Dittw \,|\, \Dittw \neq \Ditw \wedge \ditw =
  \Ditw] \\
  &= ( r_i + (1 - r_i) \cdot B)
  \end{split}
\end{equation}
In this case the agent was correct and the target function
changes. This means that if the agent retains the same mapping,
which it does with probability $r_i$, then the agent will definitely be
incorrect at time $t+1$. If it does not retain the same mapping, which
happens with probability $1-r_i$, then it will be incorrect with
probability $B$, where
\begin{equation}
  \begin{split}
  B = \Pro[ &\dittw \neq \Dittw | \ditw = \Ditw \wedge \Dittw \neq
  \Ditw \label{eq:b}\\
  & \wedge \dittw \neq \Ditw].
  \end{split}
\end{equation}
Finally, the fourth conditional probability has a value of
\begin{equation}
  \label{eq:10}
  \begin{split}
  \Pro[&\dittw \neq \Dittw \,|\, \Dittw \neq \Ditw \wedge \ditw \neq
  \Ditw] \\
  &= (1 - c_i)D + l_i +  (c_i - l_i)F,    
  \end{split}
\end{equation}
where
\begin{align}
  D &= \Pro[ \ditw \neq \Dittw | \ditw \neq \Ditw \wedge \Dittw \neq
  \Ditw] \label{eq:d}\\
  F &= \Pro[ \dittw \neq \Dittw | \ditw \neq \Ditw \wedge \Dittw \neq
  \Ditw \label{eq:f} \\
  & \quad \quad \wedge \dittw \neq \Ditw \wedge \dittw \neq \ditw]. \nonumber
\end{align}
This is the case where the target function changes and the agent was
wrong. We have to consider three possibilities. The first possibility
is for the agent not to change its decision function, which happens
with probability $1 - c_i$. The probability that the agent will be
incorrect in this case is given by $D$. The second possibility, when
the agent changes its mapping to the correct function, has a
probability of $l_i$ and ensures that the agent will be incorrect the
next time.  The third possibility happens, with probability $c_i -
l_i$ when the agent changes its mapping to an incorrect value. In this
case, the probability that it will be wrong next time is given by $F$.

We can substitute Eqs.~\eqref{eq:3}, \eqref{eq:4}, \eqref{eq:5}, and
\eqref{eq:10} into Eq.~\eqref{eq:2}, substitute the values of $a$ and
$b$, and expand $\Pro[a \wedge b]$ into $\Pro[a \,|\,b] \cdot
\Pro[b]$, in order to get
\begin{equation} 
  \label{eq:main:general}
  \begin{split}
  E&[\editt] = E[\sum_{w \in W} \Dw \Pro[\dittw \neq \Dittw]] = \sum_{w \in W} \Dw ( \\
  &\quad \Pro[\Dittw = \Ditw | \ditw = \Ditw] \cdot \Pro[\ditw = \Ditw] \cdot ( 1 -r_i) \\
  &+ \Pro[\Dittw = \Ditw | \ditw \neq \Ditw]\cdot \Pro[\ditw \neq \Ditw] \cdot (1 - l_i) \\
  &+ \Pro[\Dittw \neq \Ditw | \ditw = \Ditw ] \\
  & \quad \cdot  
    \Pro[\ditw = \Ditw] \cdot
    \left(
      r_i + (1 - r_i) \cdot B
    \right) \\
 &+ \Pro[\Dittw \neq \Ditw | \ditw \neq \Ditw ]  \cdot \Pro[\ditw \neq
 \Ditw] \\
 & \quad \cdot  (1 - c_i)D + l_i +  (c_i - l_i)F. \\
  \end{split}
\end{equation}
Equation~\eqref{eq:main:general} will model any MAS whose agent
learning can be described with the parameters presented
Section~\ref{sec:Model-Learn-Algor} and whose action/learn loop is the
same as we have described. We can use Eq.~\eqref{eq:main:general} to
calculate the successive expected errors for agent $i$, given values
for all the parameters and probabilities. In the next section we show
how this is done in a simple example game.

\subsection{The Matching game}
\label{sec:Matching-game}

In this matching game we have two agents $i$ and $j$ each of whom, in
every world $w$, wants to play the same action as the other one. Their
set of actions is $A_i = A_j$, where we assume $|A_i| > 2$ (for $|A_i|
= 2$ the equation is simpler).  After every time step, the agents both
learn and change their decision functions in accordance to their
learning rates, retention rates, and change rates.  Since the agents
are trying to match each other, in this game it is always true that
$\Delta_i^t(w)= \delta_j^t(w)$ and $\Delta_j^t(w) = \delta_i^t(w)$.
Given all this information, we can find values for some of the
probabilities in Eq.~\eqref{eq:main:general} (including values
for Equations~\eqref{eq:b} \eqref{eq:d} \eqref{eq:f}) and rewrite
(see Appendix~\ref{sec:Deriving-C-matching} for derivation) it as:
\begin{equation} 
  \label{eq:main:matching}
  \begin{split}
  E&[\editt] = \sum_{w \in W} \Dw \{  r_j \cdot \Pro[\ditw = \Ditw] \cdot ( 1 -r_i) \\
  &+ (1 -c_j) \cdot \Pro[\ditw \neq \Ditw] \cdot (1 - l_i) \\
  &+ (1-r_j) \cdot  
    \Pro[\ditw = \Ditw] \cdot
  \left(
    r_i + (1 - r_i) \cdot
    \left(
      \frac{|A_i| - 2}{|A_i| - 1}
    \right)
  \right)\\
  &+ c_j \cdot \Pro[\ditw \neq \Ditw] \cdot 
  \left(
 1 - l_j + 
    \frac{c_i l_j(|A_i| -1) + l_i(1 - l_j) - c_i}{|A_i|-2} 
  \right) \}\\
  \end{split}
\end{equation} 
We can better understand this equation by plugging in some values and
simplifying. For example, lets assume that $r_i = r_j = 1$ and $l_i =
l_j = 1$, which implies that $c_i = c_j = 1$. This is the case where
the two agents always change all their incorrect mappings so as to
match their respective target functions at time $t$. That is, if we
had $\delta_i^t(w_1)= x$ and $\delta_j^t(w_1) = y$, then at time $t+1$
we will have $\delta_i^{t+1}(w_1) = y$ and $\delta_j^{t+1}(w_1) = x$.
This means that agent $i$ changes all its incorrect mappings to match
$j$, while $j$ changes to match $i$, so all the mappings stay wrong
after all (i.e., $i$ ends up doing what $j$ did before, while $j$ does
what $i$ did before). The error, therefore, stays the same.  We can
see this by plugging the values into Eq.~\eqref{eq:main:matching}.
The first three terms will become 0 and the fourth term will simplify
to the definition of error, as given by Eq.~\eqref{eq:error}. Since
the fourth term is the only one that is non-zero, we end up with
$E[\editt] = \edit$.

We can also let $c_i$ and $l_i$ (keeping $c_j = l_j = 1$) be 
arbitrary numbers, which gives us $ E[\editt] = c_i \edit$.  This
tells us that the error will drop faster for a smaller change rate
$c_i$. The reason is that $i$'s learning (remember $l_i \leq c_i$) in
this game is counter-productive because it is always made invalid by
$j$'s learning rate of $1$. That is, since $j$ is changing all its
mappings to match $i$'s actions, $i$'s best strategy is to keep its
actions the same (i.e., $c_i = 0$).

\section{Further Simplification}
\label{sec:Assum-Cond-Indep}

We can further simplify Eq.~\eqref{eq:main:general} if we are willing to
make two assumptions. The first assumption is that the new actions
chosen when either \ditw{} changes (and does not match the target), or
when \Ditw{} changes, are both taken from flat probability
distributions over $A_i$. By making this assumption we can find
values for $B$, $D$, and $F$, namely:
\begin{alignat}{2}
  \label{eq:23}
  B = D & = \frac{|A_i| -2}{|A_i| - 1} & \qquad F & = \frac{|A_i| -3}{|A_i| - 2}
\end{alignat}


The second assumption we make is that the probability of \Ditw{}
changing, for a particular $w$, is independent of the
probability that \ditw{} was correct. In
Section~\ref{sec:Matching-game} we saw that in the matching game the
probabilities of \Ditw{} and \ditw{} changing were correlated since,
if \ditw{} was wrong then \djtw{} was also wrong, which meant $j$
would probably change \djtw{}, which would change \Ditw{}. 

However, the matching game is a degenerate example in exhibiting such
tight coupling between the agents' target functions. In general, we
can expect that there will be a number of MASs where the probability
that any two agents $i$ and $j$ are correct is uncorrelated (or
loosely correlated). For example, in a market system all sellers try
to bid what the buyer wants, so the fact that one seller bids the
correct amount says nothing about another seller's bid. Their bids are
all uncorrelated. In fact, the Distributed Artificial Intelligence
literature is full of systems that try to make the agents' decisions
as loosely-coupled as possible \cite{lesser:81,Liu:95}.

This second assumption we are trying to make can be formally
represented by having Eq.~\eqref{eq:cond:indep} be true for all pairs of
agents $i$ and $j$ in the system.
\begin{equation}
  \label{eq:cond:indep} %
  \begin{split}
  \Pro[&\ditw = \Ditw \wedge \djtw = \Djtw] \\
  & = \Pro[\ditw = \Ditw] \cdot \Pro[\djtw = \Djtw]    
  \end{split}
\end{equation}

Once we make these two assumptions we can
rewrite Eq.~\eqref{eq:main:general} as:
\begin{equation} 
  \label{eq:main}
  \begin{split}
  E&[\editt]  = \sum_{w \in W} \Dw (  \Pro[\Dittw = \Ditw] \cdot ( \Pro[\ditw = \Ditw] \cdot ( 1 -r_i) \\
  & \qquad \qquad \qquad \qquad \qquad + \Pro[\ditw \neq \Ditw] \cdot (1 - l_i)) \\
  & \quad + \Pro[\Dittw \neq \Ditw] \cdot
    (\Pro[\ditw = \Ditw] \cdot
  \left(
    r_i + (1 - r_i) \cdot
    \left(
      \frac{|A_i| - 2}{|A_i| - 1}
    \right)
  \right)\\
  & \qquad \qquad \qquad \qquad \qquad + \Pro[\ditw \neq \Ditw] \cdot 
  \left(
    \frac{|A_i| -2 - c_i + 2l_i}{|A_i| - 1}
  \right))) \\
  \end{split}
\end{equation} 
Some of the probabilities in this equation are just the definition of
$v_i$, and others simplify to the agent's error. This means that we
can simplify Eq.~\eqref{eq:main} to:
\begin{multline}
  \label{eq:main:simp}
    E[\editt] = 1 - r_i + v_i
    \left(
      \frac{|A_i|r_i - 1}{|A_i| -1}
    \right) \\
    + \edit
    \left(
      r_i -l_i + v_i
      \left(
        \frac{|A_i|(l_i - r_i) + l_i - c_i}{|A_i| -1}
      \right)
    \right)
\end{multline}

Eq.~\eqref{eq:main:simp} is a difference equation that can be used to
determine the expected error of the agent at any time by simply using
$E[\editt]$ as the \edit{} for the next iteration. While it might look
complicated, it is just the function for a line $y = mx+b$ where $x =
\edit$ and $y= \editt$. Using this observation, and the fact that
\editt{} will always be between $0$ and $1$, we can determine that the
final convergence point for the error is the point
where Eq.~\eqref{eq:main:simp} intersects the line $y=x$. The only
exception is if the slope equals $-1$, in which case we will see the
error oscillating between two points.
\begin{figure}
  \psfrag{edit}{{\small \edit}}
  \psfrag{editt}{{\small \editt}}
  \psfrag{editt2}[B]{{\tiny \editt}}
  \psfrag{learning}[B]{{\tiny learning}}
  \psfrag{volatility}[B]{{\tiny volatility}}
  \begin{center}
    \includegraphics[width=3in]{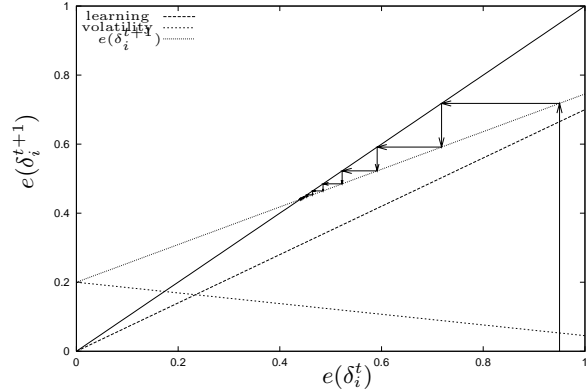}
    \caption{Error progression for agent $i$, assuming a fixed
      volatility $v_i = .2$, $c_i =1$, $l_i = .3$, $r_i = 1$, $|A_i| =
      20$. We show the error function (\editt{}), as well as its two
      components: learning and volatility. The line $y=x$ allows us
      trace the agent's error as it starts at $.95$ and converges to
      $.44$.}
    \label{fig:errorprog}
  \end{center}
\end{figure}

By looking at Eq.~\eqref{eq:main:simp} we can also determine that
there are two ``forces'' acting on the agent's error: volatility and
the agent's learning abilities. The volatility tends to increase the
agent's error past its current value while the learning reduces it. We
can better appreciate this effect by separating the $v_i$ terms in
Eq.~\eqref{eq:main:simp} and plotting the $v_i$ terms (volatility) and
the rest of the terms (learning) as two separate lines. By definition,
these will add up to the line given by Eq.~\eqref{eq:main:simp}. We
have plotted these three lines and traced a sample error progression
in Figure~\ref{fig:errorprog}. The error starts at .95 and then
decreases to eventually converge to .44.  We notice the learning curve
always tries to reduce the agent's error, as confirmed by the fact
that its line always falls below $y=x$.  Meanwhile, the volatility
adds an extra error. This extra error is bigger when the agent's error
is small since, any change in the target function is then likely to
increase the agent's error.

\section{Volatility and Impact}
\label{sec:Volatility}
Equation~\eqref{eq:main:simp} is useful for determining the agent's error
when we know the volatility of the system.  However, it is likely that
this value is not available to us (if we knew it we would already know
a lot about the dynamics of the system).  In this section we determine
the value of $v_i$ in terms of the other agents' changes in their
decision functions. That is, in terms of $\Pro[\djtt \neq \djt]$, for
all other agents $j$.

In order to do this we first need to define the \textbf{impact} \Iji{}
that agent $j$'s changes in its decision function have on $i$'s target
function.
\begin{equation}
  \label{eq:14}
 \forall_{w \in W} \;  I_{ji} = \Pro[\Dittw \neq \Ditw \,|\, \djttw \neq \djtw]
\end{equation}

We can now start to define volatility by first determining that, for
two agents $i$ and $j$
\begin{equation}
  \label{eq:15}
  \begin{split}
     \forall_{w \in W} \; v_{i}^{t} &= \Pro[\Dittw \neq \Ditw] \\
     &= \Pro[\Dittw \neq \Ditw \,|\, \djttw \neq \djtw] \cdot \Pro[\djttw
     \neq \djtw] \\
     &+ \Pro[\Dittw \neq \Ditw \,|\, \djttw = \djtw] \cdot \Pro[\djttw
     = \djtw]. \\
  \end{split}
\end{equation}

The reader should notice that volatility is no longer constant; it
varies with time (as recorded by the superscript). The first
conditional probability in Eq.~\eqref{eq:15} is just $I_{ji}$. The
second one we will set to $0$, since we are specifically interested in
MASs where the volatility arises \emph{only} as a side-effect of the
other agents' learning. That is, we assume that agent $i$'s target
function changes only when $j$'s decision function changes. For cases
with more than two agents, we similarly assume that one agent's target
function changes only when some other agent's decision function
changes.  That is, we ignore the possibility that outside influences
might change an agent's target function.

We can simplify Eq.~\eqref{eq:15} and generalize it to $N$ agents,
under the assumption that the other agents' changes in their decision
functions will not cancel each other out, making \Dit{} stay the same
as a consequence. $v_i^t$ then becomes
\begin{equation}
  \label{eq:16}
  \begin{split}
       \forall_{w \in W} \; v_{i}^{t} &= \Pro[\Dittw \neq \Ditw] \\
       &= 1 - \prod_{j \in N_{-i}}(1 - \Iji \Pro[\djttw \neq \djtw]). \\
  \end{split}
\end{equation}

We now need to determine the expected value of $\Pro[\djttw \neq
\djtw]$ for any agent. Using $i$ instead of $j$ we have
\begin{equation}
  \label{eq:55}
  \begin{split}
  \forall_{w \in W} \; \Pro[&\dittw \neq \ditw] \\
    &= \Pro{}[\ditw \neq \Ditw] \cdot \Pro[\dittw \neq \Ditw \,|\, \ditw
    \neq \Ditw] \\
    &+ \Pro{}[\ditw = \Ditw] \cdot \Pro[\dittw \neq \Ditw \,|\, \ditw
    = \Ditw], \\
  \end{split}
\end{equation}
where the expected value is:
\begin{equation}
  \label{eq:30}
  E[\Pro[\dittw \neq \ditw]] = c_i \edit + (1- r_i)\cdot(1 -
  \edit).
\end{equation}

We can then plug Eq.~\eqref{eq:30} into Eq.~\eqref{eq:16} in order to get the
expected volatility
\begin{equation}
  \label{eq:6}
 E[ v_{i}^{t}] =  1 - \prod_{j \in N_{-i}}1 - \Iji (c_j \edjt + (1- r_j)\cdot(1 -
  \edjt)).
\end{equation}

We can use this expected value of $v_{i}^{t}$ in Eq.~\eqref{eq:main:simp}
in order to find out how the other agents' learning will affect agent
$i$. In MASs that have identical learning agents (i.e., their $c$, $l$,
$r$, and $I$ rates are all the same and they start with the same
initial error) we can replace the multiplier in  Eq.~\eqref{eq:6} with an
exponent of $|N| -1$. We use this simplification later in
Section~\ref{sec:Shoham-Tennenholtz}.

\section{An Example with Two Agents}
\label{sec:Example-with-2}

In a MAS with just two agents $i$ and $j$, we can use Eq.~\eqref{eq:6} to
rewrite Eq.~\eqref{eq:main:simp} as
\begin{equation}
  \label{eq:7}
  \begin{split}
    E&[\editt] = 1 - r_i + \Iji(c_j \edjt + (1- r_j)\cdot(1 - \edjt))
    \left(
      \frac{|A_i|r_i - 1}{|A_i| -1}
    \right) \\
    &+ \edit
    \{ r_i -l_i + \Iji(c_j \edjt + (1- r_j)\cdot(1 - \edjt))
    \\
    &
      \qquad \qquad \cdot
    \left(
      \frac{|A_i|(l_i - r_i) + l_i - c_i}{|A_i| -1}
    \right)
  \}.
\end{split}
\end{equation}

\begin{figure}
  \psfrag{Iij}{{\tiny $\Iij$}}
  \psfrag{Iji}{{\tiny $\Iji$}}
  \psfrag{Error}[B]{{\tiny Error}}
  \psfrag{Final Error for i}[B]{{\small Final Error for $i$}}
  \begin{center}
    \includegraphics[width=3in]{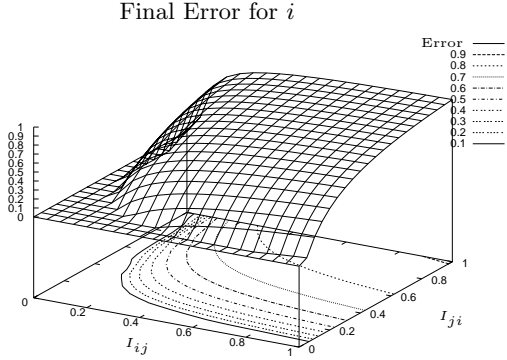}
    \caption{Plot of Final Error for agent $i$, given $l_i = l_j = .2$,
      $r_i = r_j = 1$, $c_i = c_j =1$, $|A_j| = |A_i| = 20$.}
    \label{fig:ii-error}
  \end{center}
\end{figure}

We can now use Eq.~\eqref{eq:7} to plot values for one particular example.
Let us say that $l_i = l_j =.2$, $c_i = c_j = 1$, $r_i = r_j = 1$,
$|A_j| = |A_i| = 20$ and we let the impacts $\Iij$ and $\Iji$ vary
between zero and one. Figure~\ref{fig:ii-error} shows the final error,
after convergence, for this situation. It shows an area where the
error is expected to be below $.1$, corresponding to low values for
either \Iij{}, \Iji{} or both. This area represents MASs that are
loosely coupled, i.e., one agent's change in behavior does not
significantly affect the other's target function. In these systems we
can expect that the error will eventually\footnote{Notice that we are
  not representing how long it takes for the error to converge. This
  can easily be done and is just one more of the parameters our theory
  allows us to explore.}  reach a value close to zero. We see that as
the impact increases the final error also increases, with a fairly
abrupt transition between a final error of 0 and bigger final errors.
This abrupt transition is characteristic of these types of systems
where there are tendencies for the system to either converge or
diverge, and both of them are self-enforcing behaviors. Notice also
that the graph is not symmetric---\Iij{} has more weight in
determining $i$'s final error than \Iji. This result seems
counterintuitive, until we realize that it is $j$'s error that makes
it hard for $i$ to converge to a small error. If \Iij{} is high then,
if $i$ has a large error then $j$'s error will increase, which will
make $j$ change its decision function often and make it hard for $i$
to reduce its error. If \Iij{} is low then, even if \Iji{} is high,
$j$ will probably settle down to a low error and as it does $i$ will
also be able to settle down to a low error.

If we were about to design a MAS we would try to build it so that it
lies in the area where the final error is zero. This way we can expect
all agents to eventually have the correct behavior. We note that a
substantial percentage of the research in DAI and MAS deals with
taking systems that are not inherently in this area of near-zero error
and designing protocols and rules of encounter so as to move them into
this area, as in \cite{rules:of:encounter}.

The fact that the final error is 1 for the case with $\Iij = \Iji = 1$
can seem non-intuitive to readers familiar with game theory. In game
theory there are many games, such as the ``matching game'' from
Section~\ref{sec:Matching-game}, where two agents have an impact of 1
on each other. However, it is known \cite{binmore2} that, in these
games, two learning agents will eventually converge to one of the
equilibria (if there are any), making their final error equal to 0.
This is certainly true, and it is exactly what we showed in
Section~\ref{sec:Matching-game}. The same result is not seen in
Figure~\ref{fig:ii-error} because the figure was plotted using our
simplified Equation.~\eqref{eq:main:simp}, which makes the simplifying
independence assumption given by Eq.~\eqref{eq:cond:indep}. This
assumption cannot be made in games such as the matching game because,
in these games, there is a correlation between the correctness of each
of the agents actions. Specifically, in the matching game it is always
true that both agents are either correct, or incorrect, but it is
never true that one of them is correct while the other one is
incorrect, i.e., either they matched, or they did not match.

\begin{figure}
  \psfrag{edit}{{\small \edit}}
  \psfrag{edjt}{{\small \edjt}}
  \begin{center}
    \includegraphics[width=3in]{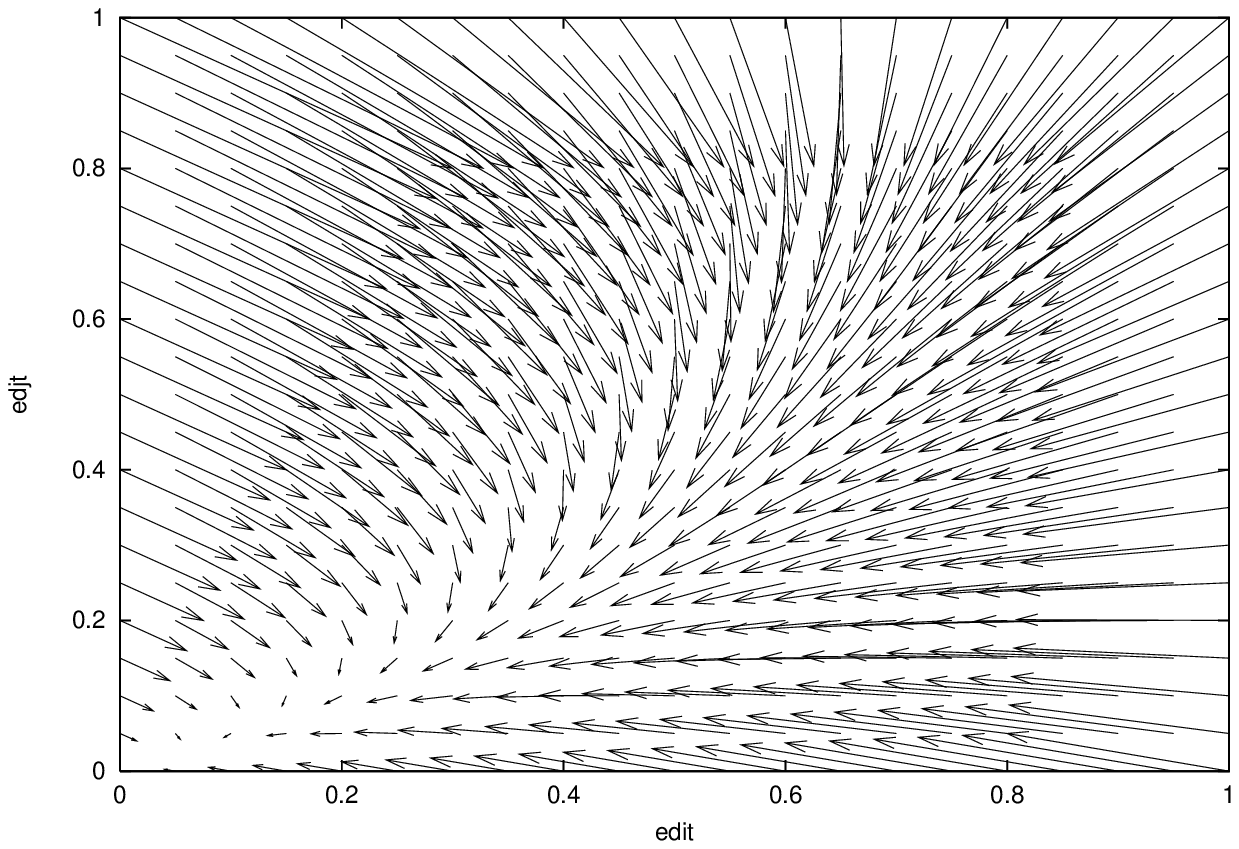}
    \caption{Vector plot for \edit{} and \edjt{}, where $|A_i| = |A_j|
      = 20$, $l_i=l_j=.2$, $r_i = r_j=1$, $c_i=.5$, $c_j=1$,
      $\Iij=.1$, $\Iji=.3$. It shows the error progression for a pair
      agents $i$ and $j$. For each pair of errors $(\edit{},\edjt{})$, 
      the arrows indicate the expected $(\editt{},\edjtt{})$.
      }
    \label{fig:vector}
  \end{center}
\end{figure}

Another view of the system is given by Figure~\ref{fig:vector} which
shows a vector plot of the agents' errors. We can see how the bigger
errors are quickly reduced but the pace of learning decreases as the
errors get closer to the convergence point. Notice also that an
agent's error need not change in a monotonic fashion. That is, an
agent's error can get bigger for a while before it starts to get
smaller.

\section{A Simple Application}
\label{sec:Simple-Application}

In order to demonstrate how our theory can be used, we tested it on a
simple market-based MAS. The game consists of three agents, one buyer
and two seller agents $i$ and $j$. The buyer will always buy at the
cheapest price---but the sellers do not know this fact. In each time
step the sellers post a price and the buyer decides which of the
sellers to buy from, namely, the one with the lowest bid.  The sellers
can bid any one of 20 prices in an effort to maximize their profits.
The sellers use a reinforcement learning algorithm with their
reinforcements being the profit the agent achieved in each round, or 0
if it did not sell the good at the time. In this system we had one
good being sold ($|W| = 1$).As predicted by economic theory, the price
in this system settles to the sellers' marginal cost, but it takes
time to get there due to the learning inefficiencies.

We experimented with different $\alpha_j$ rates\footnote{$\alpha$ is
  the relative weight the algorithm gives to the most recent payoff.
  $\alpha =1$ means that it will forget all previous experience and
  use only the latest payoff to determine what action to take.} for
the reinforcement learning of agent $j$, while keeping $\alpha_i = .1$
fixed, and plotted the running average of the error of agent $i$.
\begin{figure}
  \psfrag{l=.1=}[B]{{\tiny $\alpha_j = .1$}}
  \psfrag{l=.3=}[B]{{\tiny $\alpha_j = .3$}}
  \psfrag{l=.9=}[]{{\tiny $\alpha_j = .9$}}
  \psfrag{l=.005=}[B]{{\tiny $l_j = .005$}}
  \psfrag{l=.04=}[B]{{\tiny $l_j = .04$}}
  \psfrag{l=.055=}[]{{\tiny $l_j = .055$}}
  \psfrag{time}{{\tiny time}}
  \psfrag{error}{{\tiny error}}
  \begin{center}
    \mbox{\subfigure[Experiment]{\label{fig:lre-exp}\charts{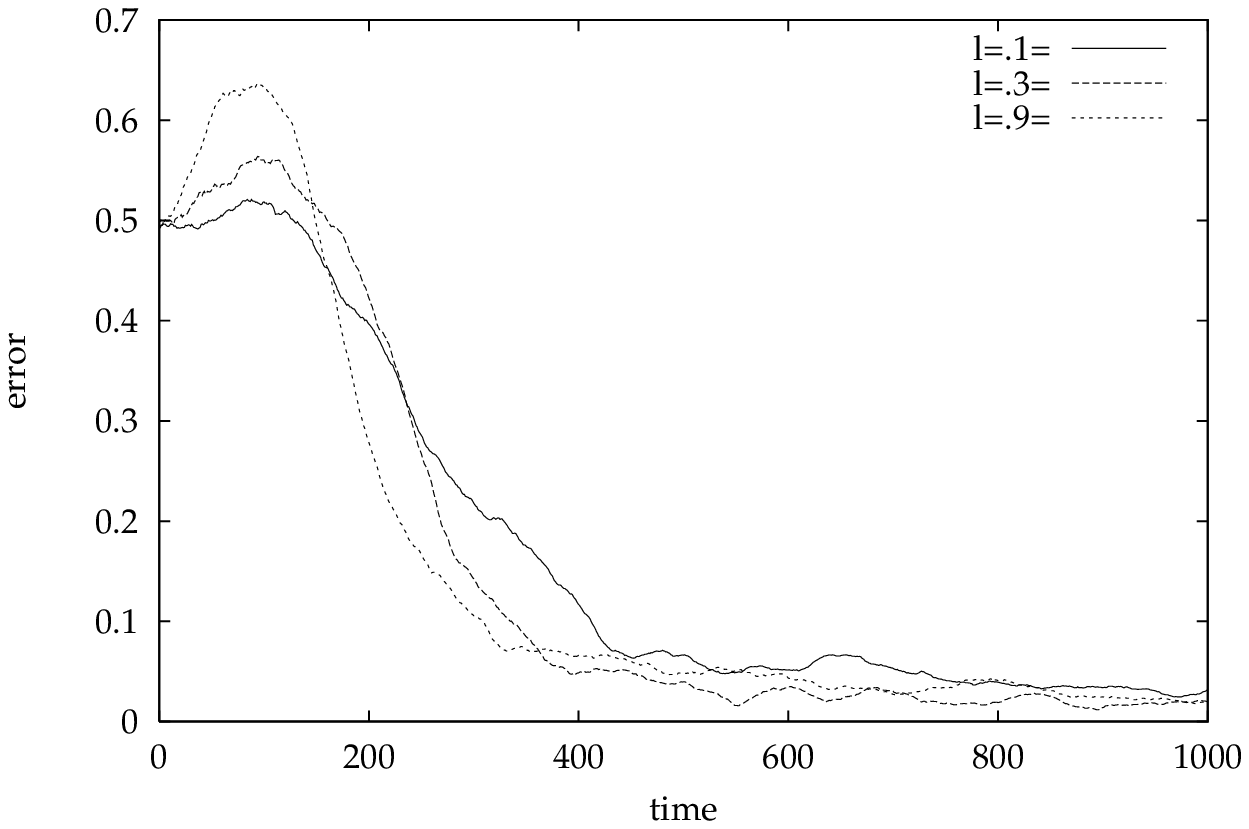}} \quad
      \subfigure[Theory]{\label{fig:lre-theory}\charts{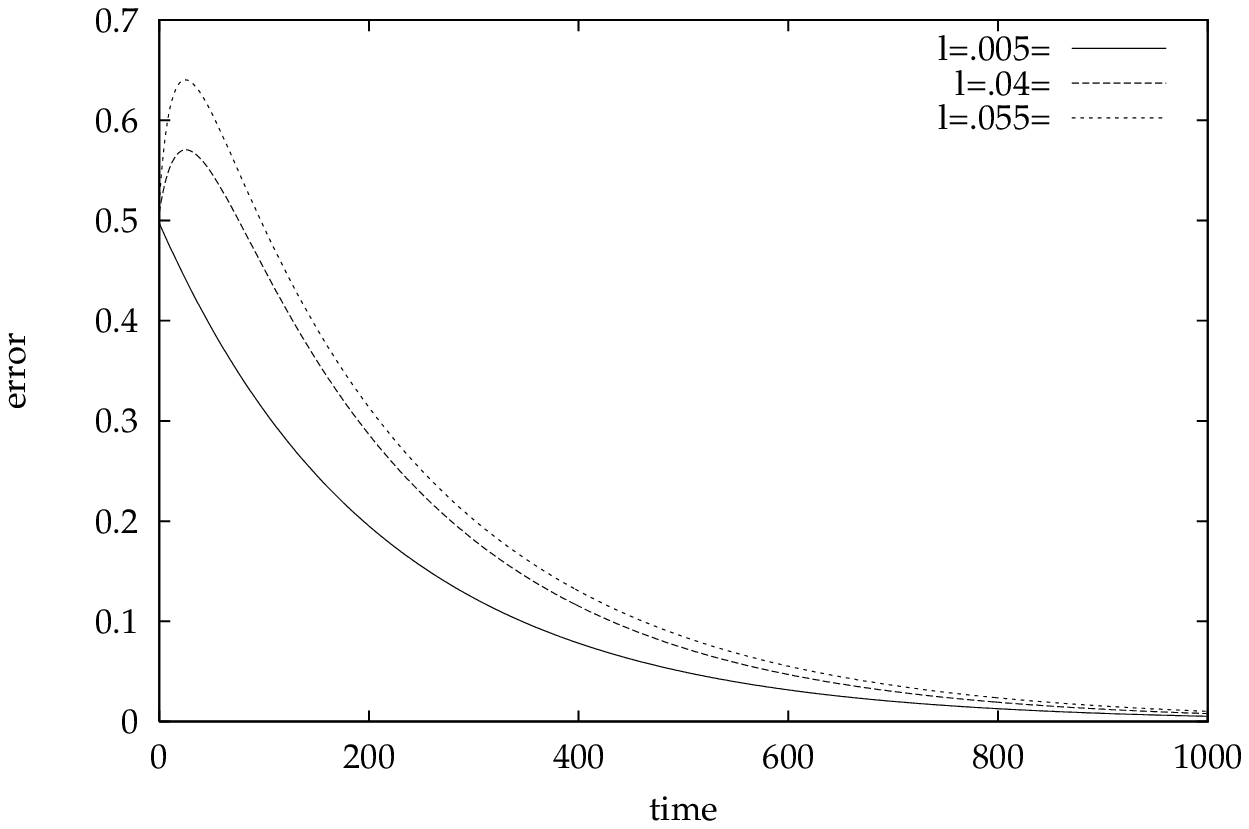}}}
    \caption{Comparison of observed and predicted error.}
    \label{fig:application}
  \end{center}
\end{figure}
A comparison is shown in Figure~\ref{fig:application}.
Figure~\ref{fig:lre-exp} gives the experimental results for three
different values of $\alpha_j$. It shows $i$'s average error, over 100
runs, as a function of time.  Since both sellers start with no
knowledge, their initial actions are completely random which makes
their error equal to $.5$. Then, depending on $\alpha_j$, $i$'s error
will either start to go down from there or will first go up some and
then down. Eventually, $i$'s error gets very close to 0, as the system 
reaches a market equilibrium.

We can predict this behavior using Eq.~\eqref{eq:7}. Based on the game
description, we set $|A_i| = |A_j| = 20$, since there were $20$
possible actions. We let $r_i = r_j = 1$ because reinforcement
learning with fixed payoffs enforces the condition that once an agent
is taking the correct action it will never change its decision
function to take a different action.  The agent might, however, still
take a wrong action but only when its exploration rate dictates it.

We then let $\Iij = \Iji = .17$ based on the rough calculation that
each agent has an equal probability of bidding any one of the 20
prices.  If $\Dit = 20$ then \Iji{} for this situation is the
probability that $j$ was also bidding 20 or above, i.e., $1/20$, times
the probability that $j$'s new price is lower than 20, i.e.  $19/20$.
Similarly, if $\Dit = 19$ then \Iji{} is equal to $2/20$ times
$18/20$.  The average of all of these probabilities is $.17$. A more
precise calculation of the impact would require us to find it via
experimentation by actually running the system.

Finally, we chose $l_i = l_j = c_i = c_j = .005$ for the first curve
(i.e., the one that compares with $\alpha_j = .1$). We knew that for
such a low $\alpha_j$ the learning and change rate should be the same.
The actual value was chosen via experimentation. The resulting curve
is shown in Figure~\ref{fig:lre-theory}. At this moment, we do not
possess a formal way of deriving learning and change rates from
$\alpha$-rates.

For the second curve ($\alpha_j = .3$) we knew that, since only
$\alpha_j$ had changed from the first experiment, we should only
change $l_j$ and $c_j$. In fact, these two values should only be
increased. We found their exact values, again by experimentation, to
be $l_j = .04$, $c_j =.4$. For the third curve we found the values to
be $l_j = .055$, $c_j = .8$.

One difference we notice between the experimental and the theoretical
results is that the experimental results show a longer delay before
the error starts to decrease. We attribute this delay to the agent's
initially high exploration rate. That is, the agents initially start
by taking all random actions but progressively reduce this rate of
exploration. As the exploration rate decreases the discrepancy between
our theoretical predictions and experimental results is reduced.

In summary, while it is true that we found $l_j$ and $c_j$ by
experimentation, all the other values were calculated from the
description of the problem. Even the relative values of $l_j$ and
$c_j$ follow the intuitive relation with $\alpha_j$ that, as
$\alpha_j$ increases so does $l_j$ and (even more) $c_j$.
Section~\ref{sec:Bound-Learn-Rate} shows how to calculate lower bounds
on the learning rate. We believe that this experiment provides solid
evidence that our theory can be used to approximately determine the
quantitative behaviors of MASs with learning agents.

\section{Application of our Theory to Experiments in the Literature}
\label{sec:Appl-our-Theory}

In this section we show how we can apply our theory to experimental
results found in the AI and MAS literature. While we will often not be
able to completely reproduce the authors' results exactly, we believe
that being able to reproduce the flavor and the main quantitative
characteristics of experimental results in the literature shows that
our theory can be widely applied and used by practitioners in this
area of research.

\subsection{Claus and Boutilier}
\begin{figure}
  \psfrag{time}{{\tiny time}}
  \psfrag{1 - error}{{\tiny 1 - error}}
  \psfrag{l=.1}[B]{{\tiny $l_i=.1$}}
  \begin{center}
    \mbox{\subfigure[Experiment]{\label{fig:claus-exp}\charts{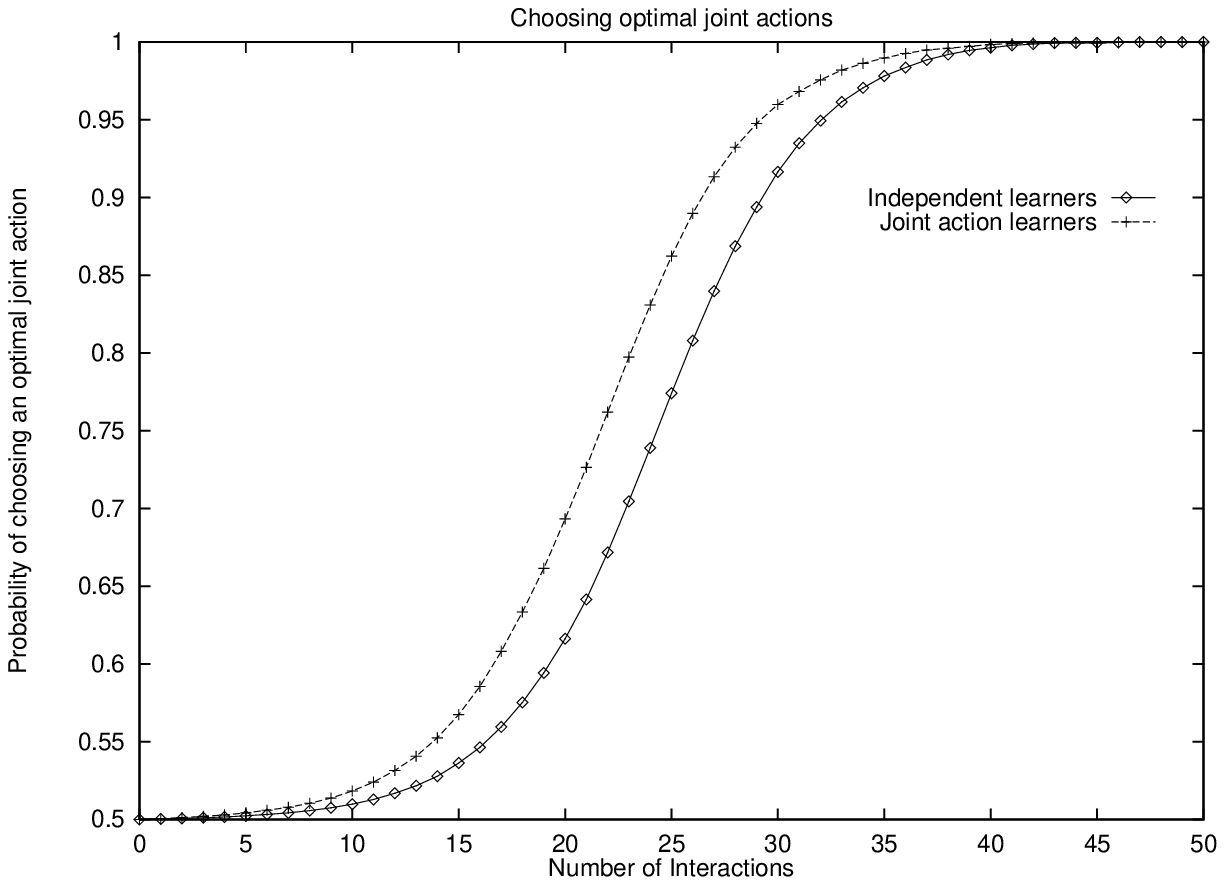}} \quad
      \subfigure[Theory]{\label{fig:claus-theory}\charts{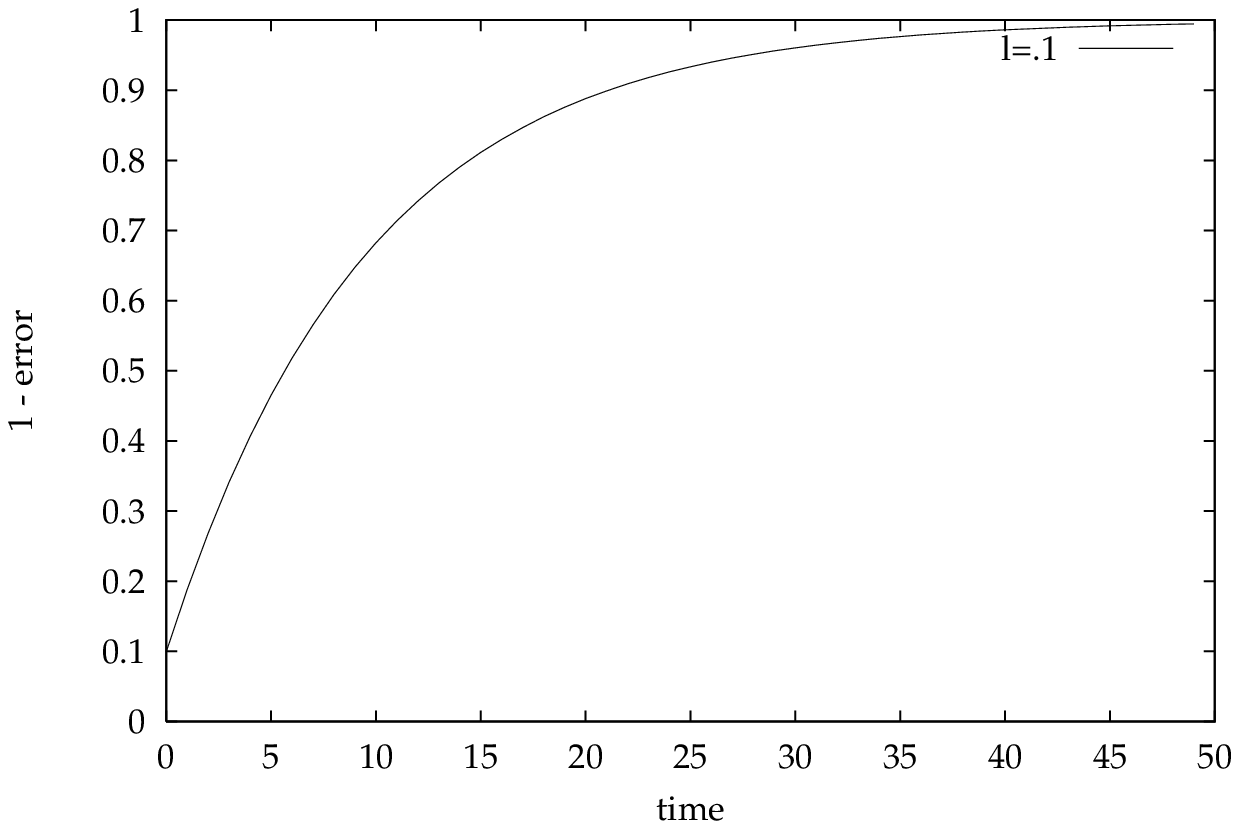}}}
    \caption{Comparing theory (b) with results from \cite{claus:97} (a).}
    \label{fig:claus}
  \end{center}
\end{figure}

Claus and Boutilier \cite{claus:97} study the dynamics of a system that
contains two reinforcement learning agents. Their first experiment
puts the two agents in a matching game exactly like the one we
describe in Section~\ref{sec:Matching-game} with $|A_i| = |A_j| = 2$.
Their results show the probability that both agents matched (i.e., 1 -
\edit) as time progressed. Since they were using two reinforcement
learning agents, it was not surprising that the curve they saw, seen
in Figure~\ref{fig:claus-exp}, was nearly identical to the curve we
saw in our experiments with the two buying agents
(Figure~\ref{fig:lre-exp} with $\alpha_j = \alpha_i = .1$, except
upside-down).

We can reproduce their curve using our equation for the matching
game Eq.~\eqref{eq:main:matching}. The results can be seen in
Figure~\ref{fig:claus-theory}. Our theory again fails to account for
the initial exploration rate. We can, however, confirm that by time 15
their Boltzmann temperature (the authors used Boltzmann exploration)
had been reduced from an initial value of 16 to $3.29$ and would keep
decreasing by a factor of .9 each time step. This means that by time
15 the agents were, indeed, starting to do more exploitation (i.e.,
reduce their error) while doing little exploration.
\label{sec:Claus-Boutilier}

\subsection{Shoham and Tennenholtz}
\label{sec:Shoham-Tennenholtz}

Shohan and Tennenholtz \cite{shoham:97} investigate how learning
agents might arrive at social conventions. The authors introduce a
simple learning algorithm (strategy-selection rule) called
\emph{highest cumulative reward} (HCR) which their agents use for
learning these conventions. Shoham and Tennenholtz also provide the
results of a series of experiments using populations of learning
agents. We try to reproduce the results they present in their
Section~4.1 where they study the ``coordination game'' which is
similar to our matching game, but with only two actions.

\begin{figure}
  \psfrag{l}{{\small $l_i$}}
  \psfrag{final error}{{\small final error}}
  \psfrag{theory}[B]{{\tiny theory}}
  \psfrag{experiment}[B]{{\tiny experiment}}
  \begin{center}
    \mbox{\subfigure[Original Experiment]{\label{fig:shoham-exp}\charts{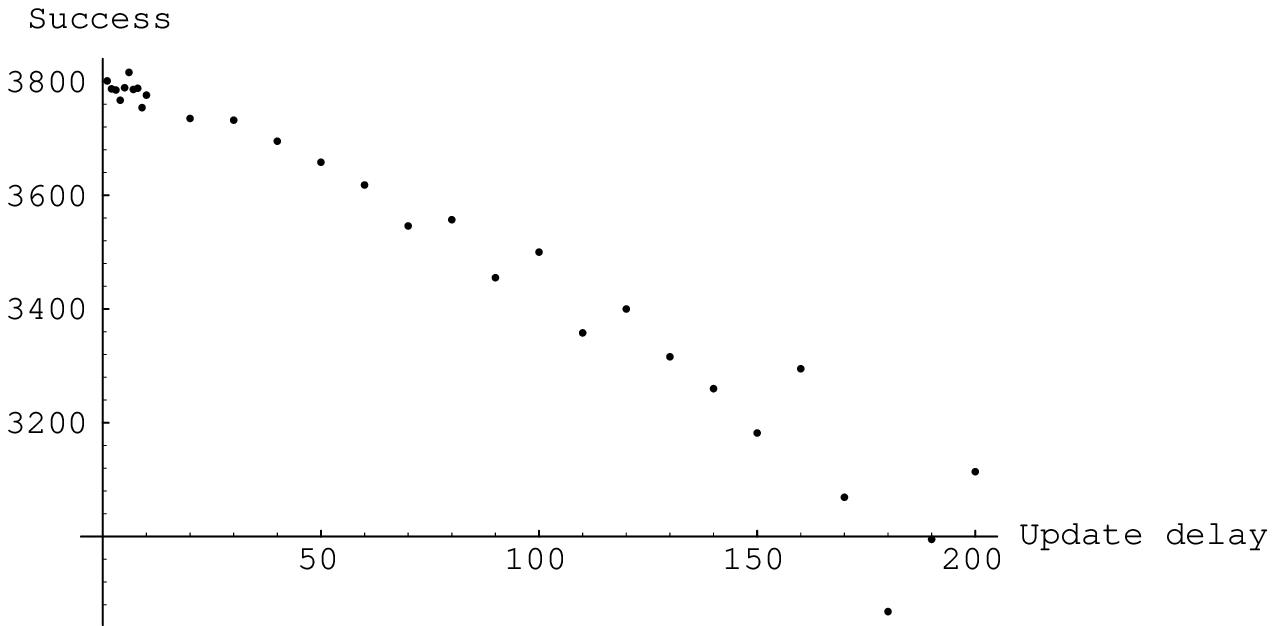}} \quad
      \subfigure[Theory and Experiment]{\label{fig:shoham-theory}\charts{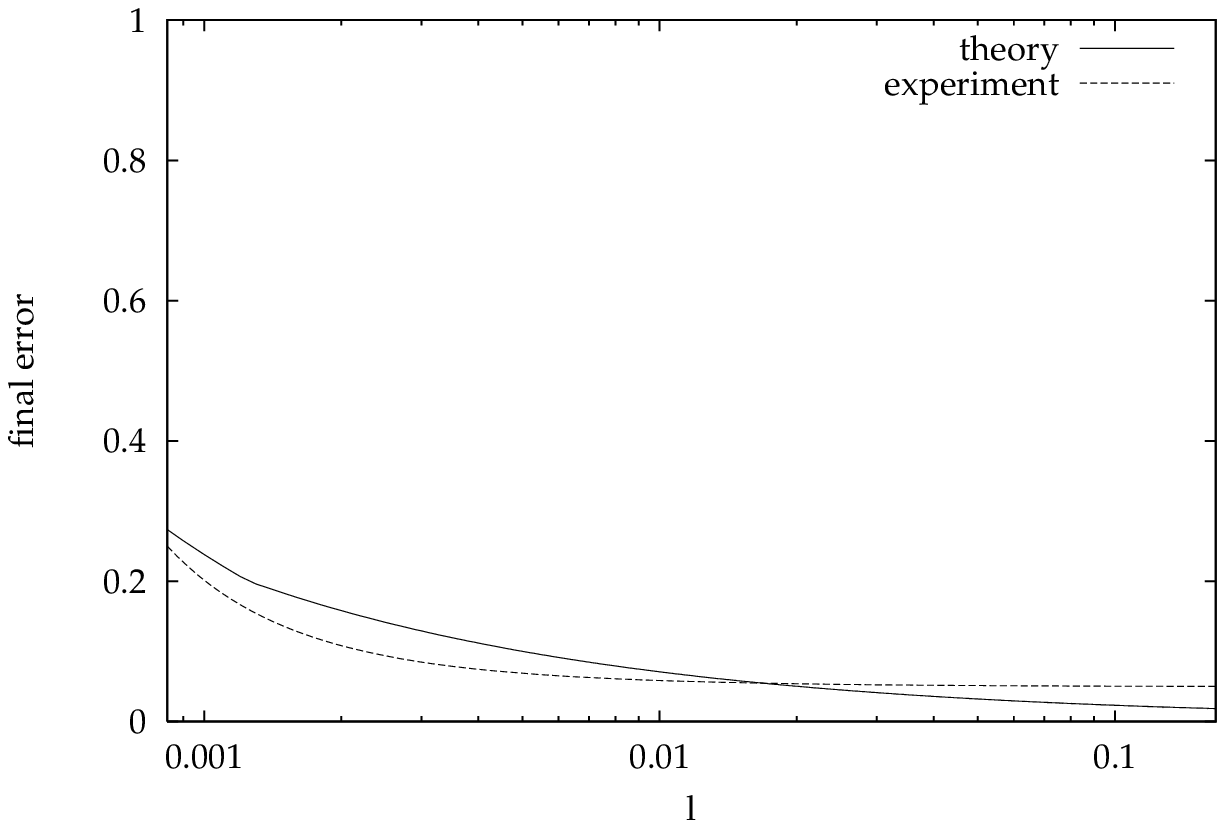}}}
    \caption{Comparing theory (b) with results from \cite{shoham:97} (a).}
    \label{fig:shoham}
  \end{center}
\end{figure}

The experiment in question involves 100 agents, all of them identical
and all of them using HCR. At each time instant the agents take one of
two available actions. The aim is for every pair of chosen agents to
take the same action as each other. Agents are randomly made to form
pairs. The agents update their behavior (i.e., apply HCR) after a
given delay. The authors try a series of delays (from 0 to 200) and
show that increasing the update delay decreases the percentage of
trials where, after 1600 iterations, at least 95\% of the agents
reached a convention. The authors show surprise at finding this
phenomenon. Their results are reproduced in
Figure~\ref{fig:shoham-exp} (cf. Figure~1 in their article).
\label{sec:Appl-to-Results}
The number of actions for all agents is easily set to $|A_i| =2$,
which implies that we must have $l_i = c_i$.  By examining HCR, it is
easy to determine that $r_i =1$ (i.e if an agent took the right
action, it will only get more support for it).  At first intuition,
one's impulse is to set $\Iij = 1$ for every pair of agents $i$ and
$j$.  However, since there are 100 of them and only pairs of them
interact at every time instant, the real impact is $\Iij = 1/99$.

We will now convert from their units of measurement into ours.  In
Figure~\ref{fig:shoham-exp} we can see that their x-axis is called the
\emph{update delay}, which we will refer to as $d$. This value is the
number of time units that pass before the agent is allowed to learn.
For $d=0$ the agent learns after every interaction (i.e., on every time
$t$), while for $d=200$ the agent takes the same action for 200 time
instances and only learns after every 200 iterations. This means that
we must set $l_i = \frac{1}{p(d +1)}$ where $p>0$. The value of $p$
depends on their learning algorithm's performance, but we know that it
must be a small number ($< 50$) greater than 0. Through some
experimentation we settled on $p=6$ (other values close to this one
give similar results). Since in their graph they look at $0 \leq d
\leq 200$, we must then look at $l_i$ where $\frac{1}{1206} \leq l_i
\leq \frac{1}{6}$ Finally, we find the value of $d$ in terms of $l_i$
to be
\begin{equation}
  \label{eq:9}
  d = \frac{1}{pl_i} - 1
\end{equation}

The y-axis of Figure~\ref{fig:shoham-exp} is the \emph{success}, i.e.,
number of trials, out of 4000, where at least 95\% of the agents
reached a convention. We will refer to this value as $s$.  We know
that in $s/4000$ of the trials \emph{at least} 95\% of the
agents have error close to 0 (i.e., reaching a convention means that
the agents take the right action almost all the time), and for the
rest of the trials the error was greater.  We can approximately map
this to an error by saying that in $s/4000$ of the trials the error
was 0 (a slight underestimate), while in $1 - s/4000$ of the trials
the error was 1 (a slight overestimate).  We add these two up (the 0
makes the first term disappear) and arrive at an equation that maps
$s$ to \edit.
\begin{equation}
  \label{eq:26}
  \edit \approx
  \left(
    \frac{4000 - s}{4000}
  \right)
\end{equation}

The mapping from $d$ to $s$ is given by their actual data. Their data
can be fit by the following function:
\begin{equation}
  \label{eq:28}
  s = 3900 - 4d - \frac{(d -100)^2}{100}
\end{equation}

Plugging Eq.~\eqref{eq:9} into Eq.~\eqref{eq:28}, and the result
into Eq.~\eqref{eq:26}, we finally arrive at a function that maps their
experimental results into our units:
\begin{equation}
  \label{eq:29}
  \mbox{Final error} = \frac{4000 -
    \left(
      3900 - 4(1/pl_i - 1) -
      \left(
        \frac{(1/pl_i - 1 -100)^2}{100}
      \right)
    \right)}{4000}
\end{equation}
for the range $\frac{1}{1206} \leq l_i \leq \frac{1}{6}$.

Now that we have values for $c_i$, $l_i$, $r_i$, \Iij, $|A_i|$, a
range for $l_i$ and an equation that maps their experimental results
into our units, we can plot both functions, as seen in
Figure~\ref{fig:shoham-theory}. The x-axis was plotted on a log-scale
in order to better show the shape of the experiment curve, otherwise
it would appear mostly as a straight line. For our theory curve we
used Equations~\eqref{eq:main:simp} and \eqref{eq:6}, and iterated for 1600 time
units, just like in the experiment, and plotted the error at that
point. For the experiment curve we used Eq.~\eqref{eq:29}. We plotted both
of these curves in the specified range for $l_i$.  The reader will
notice that our theory was able to make precise quantitative
predictions. The maximum distance from our theory curve to the
experimental curve is $.05$, which means that our predictions for the
final error were, at worst, within 5\% of the experimental values.
Also, an error of about 5\% was introduced when mapping from their
success percentage $s$ to our error.

\subsection{Others}
\label{sec:Others}

There are several other examples in the literature where we believe
our theory can be successfully applied. \cite[chapter 3.7]{ishida:97}
gives results of an experiment where two agents try to find each other
in a 100 by 100 grid. He shows that if the grid has few obstacles it
is faster if both agents move towards each other, while if there are
many obstacles it is faster if one of the agents stays still while the
other one searches for it. We believe that the number of obstacles is
proportional to the change rate that the agents experience and,
perhaps, to the impact that they have on each other.  When there are
no obstacles the agents never change their decision functions (because
their initial Manhattan heuristics lead them in the correct path). As
the number of obstacles increases, the agents will start to change
their decision functions as they move, which will have an impact on
the other agent's target function.  If, however, one of them stays
put, this means that his change rate is 0 so the other agent's target
function will stay still and he will be able to reach his target
(i.e., error 0) quicker.

Notice that the problem of a moving target that Ishida studies is
different from the problem of a moving target function which we study.
It is, however, interesting to note their similarities and how our
theory can be applied to some aspects of that domain.

Another possible example is given by \cite{sen:94b}.  They show two
Q-learning agents trying to cooperate in order to move a block.  The
authors show how different $\alpha$ rates ($\beta$ in their article)
affect the quality of the result that the agents converge to. This
quality roughly corresponds to our error, except for the fact that
their measurements implicitly consider some actions to be better than
others, while we consider an action to be either correct or incorrect.
This discrepancy would make it harder to apply our theory to their
results but we still believe that a rough approximation is possible.
Our future work includes the extension of the CLRI framework to handle
a more general definition of error---one that attaches a utility to
each state-action pair, rather than the simple correct/incorrect
categorization we use.

\section{Bounding the Learning Rate with Sample Complexity}
\label{sec:Bound-Learn-Rate}

In the previous examples we have used our knowledge of the learning
algorithms to determine the values of the agent's $c_i$, $l_i$, and
$r_i$ parameters. However, there might be cases where this is not
possible---the learning algorithm might be too complicated or unknown.
It would be useful, in these cases, to have some other measure of the
agent's learning abilities, which could be used to determine some
bounds on the values of these parameters.

One popular measure of the complexity of learning is given by Probably
Approximately Correct (PAC) theory \cite{intro:clt}, in the form of a
measure called the \emph{sample complexity}. The sample complexity
gives us a loose upper bound on the number of examples that a
consistent learning agent must observe before arriving at a PAC
hypothesis.

There are two important assumptions made by PAC-theory. The first
assumption is that the agents are consistent learners\footnote{See
  \cite[p162]{machine:learning} for a formal definition of a
  consistent learner.}. Using our notation, a consistent learner is one
who, once it has learned a correct $w \ra a$ mapping does not forget
it. This simply means that the agent must have $r_i = 1$. The second
assumption is that the agent is trying to learn a fixed concept. This
assumption makes $\Ditt = \Dit$ true for all $t$.

The sample complexity $m$ of an agent's learning problem is given by
\begin{equation}
  \label{eq:11}
  m \geq \frac{1}{\epsilon}
  \left(
    \ln \frac{|H|}{\gamma}
  \right),
\end{equation}
where $|H|$ is the size of the hypothesis space for the agent. In
other words, $|H|$ is the total number of different \diw{} functions
that the agent will consider. For an agent with no previous knowledge
we have $|H| = |A_i|^{|W|}$. However, agents with previous knowledge
might have smaller $|H|$, since this knowledge might be used to
eliminate impossible mappings. If a consistent learning agent has seen 
$m$ examples then, with probability at least $(1 - \gamma)$, it has
error at most $\epsilon$.

While we cannot map the sample complexity $m$ to a particular learning
rate $l_i$, we can use it to put a lower bound on the learning rate
for a consistent learning agent. That is, we can find a lower bound
for the learning rate of an agent who does not forget anything it has
seen, and who is trying to learn a fixed target function.  Since the
agent does not forget anything it has seen, we can deduce that its
retention rate must be $r_i = 1$. Since the target function is not
changing, we know that $\Pr[\Dittw \neq \Ditw] = 0$ and $\Pr[\Dittw =
\Ditw] = 1$. We can plug these values into Eq.~\eqref{eq:main} and
simplify in order to get:
\begin{equation}
  \label{eq:13}
  E[\editt] = \edit \cdot (1 - l_i).
\end{equation}
We can solve the difference Eq.~\eqref{eq:13}, for any time $n$, in
order to get:
\begin{equation}
  \label{eq:17}
  E[e(\delta_i^n)] = e(\delta_i^0) \cdot (1 - l_i)^n.
\end{equation}
We now remember that after $m$ time steps we expect, with probability
$(1 - \gamma)$, the error to be less than $\epsilon$.  Since
Eq.~\eqref{eq:17} only gives us an expected error, not a probability
distribution over errors, we cannot use it to calculate the likelihood
of the agent having that expected error. That is, we cannot calculate
the ``probably'' ($\gamma$) part of probably approximately correct. We
will, therefore, assume that the $\gamma$ chosen for $m$ is small
enough so that it will be safe to say that, after $m$ time steps, the
error is less than $\epsilon$.  In a typical application one uses a
small $\gamma$ because it guarantees a high degree of certainty on the
upper bound of the error.

Since we can now safely say that, after $m$ time steps, the error is
less than $\epsilon$, we can then deduce that the $l_i$ for this agent
should be small enough such that, if $n = m$, then $E[e(\delta_i^n)]
\leq \epsilon$.  This is expressed mathematically as:
\begin{equation}
  \label{eq:18}
  e(\delta_i^0) \cdot (1 - l_i)^m \leq \epsilon.
\end{equation}

We solve this equation for $l_i$ in order to get:
\begin{equation}
  \label{eq:19}
  l_i \geq 1 -
  \left(
    \frac{\epsilon}{e(\delta_i^0)}
  \right)^{1/m}.
\end{equation}
This equation is not defined for $e(\delta_i^0) = 0$. However, given
our assumption of a fixed target function and $r_i = 1$, we already
know, from Eq.~\eqref{eq:13}, that if an agent starts with an error of
$0$ it will maintain this error of $0$ for any future time $t > 0$.
Therefore, in this case, the choice of a learning rate has no bearing
on the agent's error, which will always be $0$.

Equation~\eqref{eq:19} gives us a lower bound on the learning rate
that a consistent learner must have, given that it has sample
complexity $m$, and based on an error $\epsilon$ and a sufficiently
small $\gamma$. A designer of an agent that uses a reasonable learning
algorithm can expect that, if his agent has sample complexity $m$ (for
$\epsilon$ error), then his agent will have a learning rate of at
least $l_i$, as given by Eq.~\eqref{eq:19}. Furthermore, if a designer
is comparing two possible agent designs, each with a different sample
complexity but both with similarly powerful learning algorithms, he
can calculate bounds on the learning rates of both agents and
compare their relative performance.

\section{Related Work}
\label{sec:Related-Work}

The topic of agents learning about agents arises often in the studies
of complexity \cite{complexity}. In fact, systems where the agents try
to adapt to endogenously created dynamics are being widely studied
\cite{hubler94, arthur97b}. In these systems, like in ours, the agents
co-create their expectations as they learn and change their behaviors.
Complexity research uses simulated agents in an effort to understand
the complex behaviors of these systems as observed in the real world.

One example is the work of Arthur \emph{et. al.} \cite{arthur97b}, who
arrive at the conclusion that systems of adaptive agents, where the
agents are allowed to change the complexity of their learning
algorithms, end up in one of two regimes: a stable/simple regime where
it is trivial to predict an agent's future behavior, and a complex
regime where the agents' behaviors are very complex. It is this second
regime that interests complexity researchers the most. In it, the
agents are able to reach some kind of ``equilibrium'' point in model
building complexity.  These same results are echoed by Darley and
Kauffman \cite{darley97a} in a similar experiment. In this article we
have not allowed the agents to dynamically change the complexity of
their learning algorithms.  Therefore, our dynamics are simpler.
Allowing the agents to change their complexity amounts to allowing
them to change the values of their $c$, $l$, and $r$ parameters while
learning.

However, while complexity research is very important and inspiring, it
is only partially relevant to our work. Our emphasis is on finding
ways to predict the behavior of MASs composed of machine-learning
agents.  We are only concerned with the behavior of simpler artificial
programmable agents, rather than the complex behavior of humans or the
unpredictable behavior of animals.

The dynamics of MASs have also been studied by Kephart \emph{et.  al.}
\cite{kephart:90}.  In this work the authors show how simple
predictive agents can lead to globally cyclic or chaotic behaviors. As
the authors explain, the chaotic behaviors were a result of the simple
predictive strategies used by the agents. Unlike our agents, most of
their agents are not engaged in learning, instead they use simple
fixed predictive strategies, such as ``if the state of the world was
$x$ ten time units before, then it will be $x$ next time so take
action $a$''. The authors later show how learning can be used to
eliminate these chaotic global fluctuations.

Matari\'{c} \cite{mataric97a} has studied reinforcement learning in
multi-robot domains. She notes, for example, how learning can give
rise to social behaviors \cite{mataric97b}. The work shows how robots
can be individually programmed to produce certain group behaviors. It
represents a good example of the usefulness and flexibility of
learning agents in multi-agent domains. However, the author does not
offer a mathematical justification for the chosen individual learning
algorithms, nor does she explain why the agents were able to converge
to the global behaviors. Our research hopes to provide the first steps
in this direction.

One particularly interesting approach is taken by Carmel and
Markovitch \cite{carmel97}.  They work on model-based learning, that
is, agents build models of other agents via observations. They use
models based on finite state machines.  The authors show how some of
these models can be effectively learned via observation of the other
agent's actions. The authors concentrate on the development of
learning algorithms that would let one agent learn a finite-state
machine model of another agent. They have not considered the case
where two or more agents are simultaneously learning about each other,
which we study in this article. However, their work is more general in
the sense that they model agents as state machines, rather than the
state-action pairs we use.

Finally, a lot of experimental work has been done in the area of
agents learning about agents \cite{acl:96, weib:97}.  For example, Sen
and Sekaran \cite{sen:98a} show how learning agents in simple MAS
converge to system-wide optimal behavior. Their agents use Q-learning
or modified classifier systems in order to learn. The authors
implement these agents and compare the performance of the different
learning algorithms for developing agent coordination. Hu and Wellman
\cite{hu:98a, hu:96} have studied reinforcement learning in
market-base MASs, showing how certain initial learning biases can be
self-fulfilling, and how learning can be useful but is affected by an
agent's models of other agents.  Claus and Boutilier \cite{claus:97}
have also carried out experimental studies of the behavior of
reinforcement learning agents.  We have been able to use the CLRI
framework to predict some of their experimental results
\cite{vidal:thesis}. Other researchers \cite{stone99a, littman94a,
  hu:98b} have extended the basic Q-learning \cite{watkins:92}
algorithm for use with MASs in an effort to either improve or prove
convergence to the optimal behavior.

We have also successfully experimented with reinforcement learning
simulations \cite{vidal:98b}, but we believe that the formal
treatment elucidated in these pages will shed more light into the real
nature of the problem and the relative importance of the various
parameters that describe the capabilities of an agent's learning
algorithm.

\section{Limitations and Future Work}
\label{sec:future-work}

The CLRI framework places some constraints on the type of systems it
can model, which limits its usability. However, it is important to
understand that, as we remove the limitations from the CLRI framework,
the dynamics of the system become much harder to predict.  In the
extreme, without any limitations on the agents' abilities, the system
becomes a complex adaptive system, as studied by Holland
\cite{holland95a} and others in the field of complexity.  The dynamic
behavior of these systems continues to be studied by complexity
researchers with only modest progress.  It is only by placing
limitations on the system that we were able to predict the expected
error of agents in the systems modeled by the CLRI framework.

Our ongoing work involves the relaxation of some of the constraints
made by the CLRI framework so that it may become more easily and
widely applicable, without making the system dynamics impossible to
analyze. We are targeting three specific constraints.
\begin{enumerate}
\item The values of $c_i$, $l_i$, $r_i$, and $I_{ij}$ cannot, in all
  situations, be mathematically determined from the system's
  description. We have found that bounds for the $c_i$, $l_i$, and
  $r_i$ values can often be determined when using reinforcement
  learning or supervised learning. However, the bounds are often very
  loose. The values of the $I_{ij}$ parameter depend on the particular
  system. Sometimes it is trivial to calculate the impact, sometimes
  it requires extensive simulation.
\item The CLRI framework assumes that an agent's action is either
  correct or incorrect. The framework does not allow degrees of
  correctness.  Specifically, in many systems the agents can often
  take several actions, any one of which is equally good. When
  modeling these systems, the CLRI framework requires the user to
  designate one of those actions as the correct one, thereby ignoring
  some possibly useful information.
\item The world states are taken from a uniform probability
  distribution which does not change over time. The environment is
  assumed to be episodic. As such, the framework is limited in the
  type of domains it can effectively describe.
\end{enumerate}

We are attacking these challenges with some of the same tools used by
researchers in complex adaptive systems, namely, agent-based
simulations and co-evolving utility landscapes. We believe we can gain
some insight into the dynamics of adaptive MASs by constructing and
analyzing various types of MASs. We also believe that the next step
for the CLRI framework is the replacement of the current error
definition with a utility function. The agents can then be seen as
searching for the maximum value in the changing utility landscape
defined by their utility function. The degree to which the agents are
successful on their climb to the landscape peaks depends on the
abilities of their learning algorithm (change rate, learning rate, and
retention rate), and the speed at which the landscape changes as the
other agents change their behavior (impact).

The use of utility landscapes will allow us to consider an agent's
utility for any particular action, rather than simply considering
whether an action is correct or incorrect. The landscapes will also
allow us to consider systems where agents cannot travel between any
two world states in one time step. That is, the agents' moves on the
landscape will be constrained in the same manner as their actions or
behaviors are constrained the actual system. Finally, the new theory
will likely need to redefine the CLRI parameters. We hope the new
parameters will be easy to derive directly from the values that govern
the machine-learning algorithms' behavior. These extensions will make
the new theory applicable to a much wider set of domains.

\section{Summary}
\label{sec:Summary}

We have presented a framework for studying and predicting the behavior
of MASs composed of learning agents. We believe that this framework
captures the most important parameters that describe an agents'
learning and the system's rules of encounter. Various comparisons
between the framework's predictions and experimental results were
given.  These comparisons showed that the theoretical predictions
closely match our experimental results and the experimental results
published by others.  Our success in reproducing these results allows
us to confidently state the effectiveness and accuracy of our theory
in predicting the expected error of machine learning agents in MASs.

Since our theory describes an agent's behavior at a high-level (i.e.,
the agent's error), it is not capable of making system-specific
predictions (e.g., predicting the particular actions that are
favored). These types of system-specific predictions can only be
arrived at by the traditional method of implementing populations of
such agents and testing their behaviors. However, we expect that there
will be times when the predictions from our theory will be enough to
answer a designer's questions. A MAS designer that only needs to
determine how ``good'' the agent's behavior will be could probably use
the CLRI framework. A designer that needs to know which particular
emergent behaviors will be favored by his agents will need to
implement the agents.

Finally, while we have given some examples as to how learning rates
can be determined for particular machine learning implementations, we
do not have any general method for determining these rates. However,
we showed how to use the sample complexity of a learning problem to
determine a lower bound on the learning rate of a consistent learning
agent. This bound is useful for quickly ruling out the possibility of
having agents with high expected errors and of stating that an agent's
expected error will be, at most, a certain constant value. Still, if
the agent's learning algorithm is much better than the one assumed by
a consistent learner (e.g., the agent is very good at generalizing
from one world state to many others), then these lower bounds could be
significantly inaccurate.

\appendix{}

\section{Derivation for Matching Game}
\label{sec:Deriving-C-matching}

If we can assume that the action chosen when an agent changes \ditw{}
and the result does not match \Ditw{} (for some specific $w$) is taken
from a flat probability distribution, then we can say that:

\begin{equation}
  \label{eq:34}
  B = \frac{|A_i| - 2}{|A_i| - 1}.
\end{equation}

We will now show how to calculate the fourth term
in~\eqref{eq:main:matching}.  For the matching game we find that we
can set:
\begin{align}
  D &= 1 - l_j \\
  F &= l_j + (1 - l_j)
  \left(
    \frac{|A_i| - 3}{|A_i| - 2}
  \right).
\end{align}
Having $|A_i| =2$ implies that $c_i = l_i$, this means
that for this case we have
\begin{equation}
  \label{eq:8}
  (1 -c_i) D + l_i + (c_i - l_i)F    = l_i + (1 -c_i)( 1 - l_j).
\end{equation}

For the case where $|A_i|> 2$, which is the case we are interested in,
we can plug in the values for $D$ and $F$ and simplify, in order to
get the fourth term:
\begin{equation}
  \label{eq:33}
  (1 -c_i) D + l_i + (c_i - l_i)F  = 1 - l_j + 
    \frac{c_i l_j(|A_i| -1) + l_i(1 - l_j) - c_i}{|A_i|-2}.
\end{equation}

\bibliographystyle{plain}
\bibliography{../annobib,../vidal}

\end{document}